\documentclass[runningheads]{svmult}

\usepackage{makeidx}   % allows index generation
\usepackage{graphicx}  % standard LaTeX graphics tool
\usepackage{subeqnar}  % subnumbers individual equations
\usepackage{multicol}  % used for the two-column index
\usepackage{physprbb}  % modified textarea for proceedings,
\makeindex             % used for the subject index
\begin{document}
\title*{Mapping the peculiar binary GP~Com}
\toctitle{Mapping the peculiar binary GP~Com}
% allows explicit linebreak for the table of content
%
%
\titlerunning{Mapping the peculiar binary GP~Com}
% allows abbreviation of title, if the full title is too long
% to fit in the running head
%
\author{L. Morales-Rueda\inst{}
\and T. R. Marsh\inst{}
\and R. C. North\inst{}}
\authorrunning{L. Morales-Rueda et al.}
% if there are more than two authors,
% please abbreviate author list for running head
%
%
\institute{University of Southampton, Southampton SO17 1BJ, UK}

\maketitle              % typesets the title of the contribution

\begin{abstract}
  We present high resolution spectra of the AM~CVn helium binary
  GP~Com at two different wavelength ranges. The spectra show the same
  flaring behaviour observed in previous UV and optical data. We find
  that the central spike contributes to the flare spectra indicating
  that its origin is probably the compact object. We also detect that
  the central spike moves with orbital phase following an S-wave
  pattern.  The radial velocity semiamplitude of the S-wave is
  $\sim$10\,km s$^{-1}$ which indicates its origin is near the centre
  of mass of the system, which in this case lies very close to the
  white dwarf. The Stark effect seems to affect significantly the
  central spike of some of the lines suggesting that it forms in a
  high electron density region. This again favours the idea that the
  central spike originates in the white dwarf. We present Doppler maps
  obtained for the emission lines which show three clear emission
  regions.

\end{abstract}

\section{Helium rich binaries}
GP~Com belongs to a group of binaries called AM~CVn systems. In these
binaries, a white dwarf accretes material from the stripped-down core
of a giant star. Only 6 known systems belong to this group although
they are predicted to have a space density about a factor of 2 higher
than that of cataclysmic variables (CVs) of which some 700 are known
\cite{ty96}. They show properties similar to those of CVs but with
some peculiarities, for instance ultra short orbital periods --
between 15 and 46\,min -- which indicate that the two stars are very
close together, and a complete lack of hydrogen in their spectra. Some
of the systems show strong flickering indicating the presence of mass
transfer.  AM~CVn and EC\,15330 show only absorption lines. These
lines have large breadths indicating either rapid rotation or pressure
broadening by compact stars. CR~Boo, CP~Eri and V803~Cen show
absorption lines when they are bright and emission lines when faint.
This absorption/emission line behaviour has been compared to that of
CVs, as most of them show emission lines when in quiescence and
absorption lines when in outburst.  GP~Com only shows emission lines
and therefore could be considered as always being in quiescence. For a
summary of the properties of AM~CVn systems see \cite{w95}.

\section{GP~Com's spectra}
We took spectra of GP~Com in two wavelength ranges,
$\lambda\lambda$6600 -- 7408\AA\ and $\lambda\lambda$4253 -- 5058\AA\ 
with the 4.2\,m William Herschel telescope during two consecutive
nights. For details on the observations and their reduction see
\cite{mmwn00}.

The average spectrum of GP~Com consists of strong emission lines,
mainly He\,{\sc i}, on a low continuum, see the top panels of
Figs.~\ref{eps1} and~\ref{eps2}.  The spectra look very similar to
previously published spectra \cite{nrs81,m99} but the intensity of the
central spike is stronger indicating that GP~Com shows long term
variability. The emission lines consist of a double-peaked profile
superposed with a narrow line component that moves between the red and
blue peaks with orbital phase, and a central narrow spike near the
rest wavelength.  The central spike seems to be independent of the
double peaks, which are associated with an accretion disc around the
compact object \cite{s75}. The narrow line component probably has its
origin in the region of impact between the accretion stream and the
accretion disc, i.e. the bright spot \cite{nrs81}.  The origin of the
central spike has been a puzzle for a long time. It was suggested that
it came from a surrounding nebula \cite{nrs81} but subsequent searches
were unsuccessful \cite{s83}. The central spike seems to participate
in the flaring activity shown by GP~Com which would suggest that it is
associated to the compact object \cite{m99}.  We present results that
encourage us to suggest that its origin is the compact object in the
binary and not a surrounding nebula.

\subsection{Flares}

\begin{figure}[b]
\begin{center}
\includegraphics[angle=90, width=.92\textwidth]{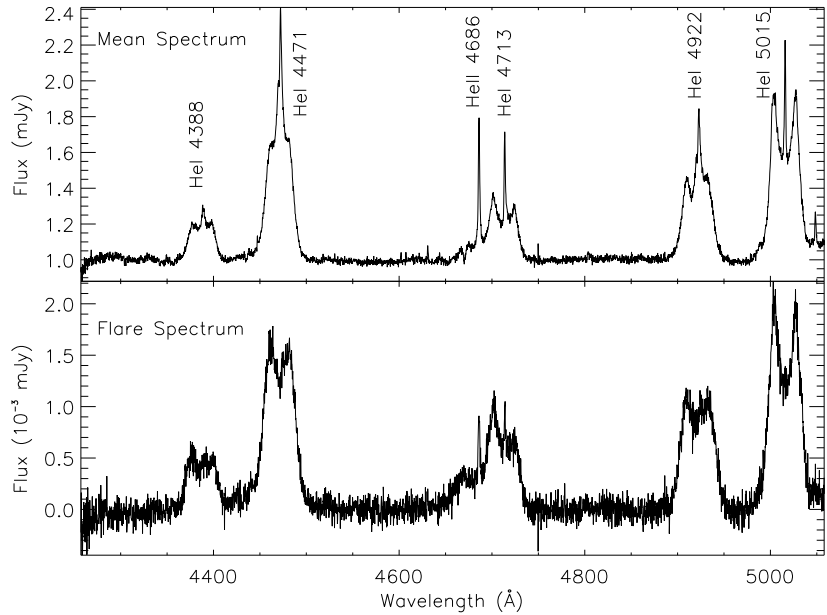}
\end{center}
\caption[]{The panels show the mean GP~Com normalised spectrum ({\bf top})
  and the flare spectrum ({\bf bottom})}
\label{eps1}
\end{figure}

\begin{figure}[t]
\begin{center}
\includegraphics[angle=90, width=.92\textwidth]{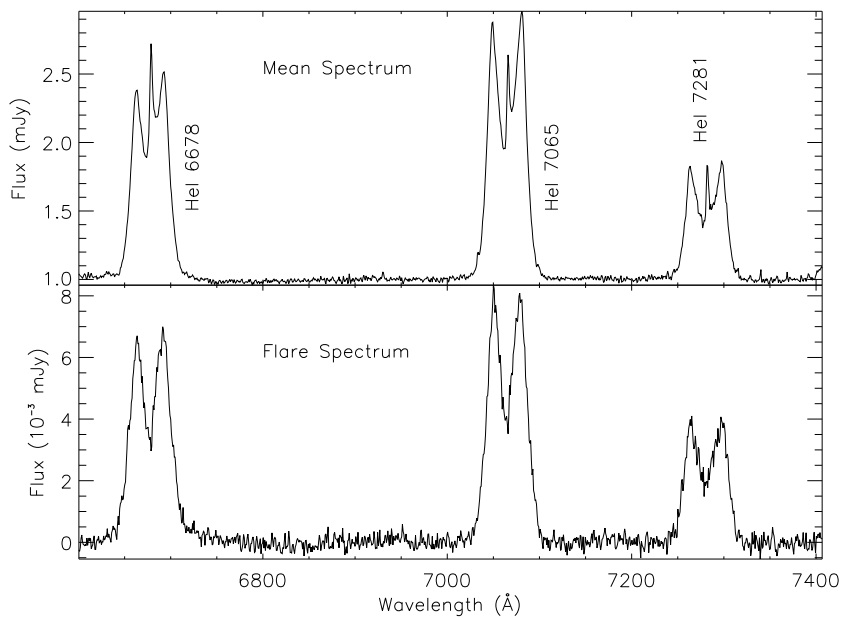}
\end{center}
\caption[]{Same as for Fig.~\ref{eps1} but for a different wavelength range}
\label{eps2}
\end{figure}

GP~Com shows strong flaring activity in UV \cite{mwhl95} and optical
\cite{m99} wavelengths probably driven by X-ray variability. By using
the method described in \cite{m99} we obtain the characteristic flare
spectra of GP~Com for both wavelength ranges, see bottom panels of
Figs.~\ref{eps1} and~\ref{eps2}. The first thing we notice is that the
central spike is not present in the flare spectra for most emission
lines but it is strong for \hbox{$\hbox{He\,{\sc
      ii}\,$\lambda$4686\,\AA}$}\ and present, although weak, for
\hbox{$\hbox{He\,{\sc i}\,$\lambda$4713\,\AA}$}. We are certain that
the central spike contributes to the flare spectrum of GP~Com and
therefore suggest that it must have its origin somewhere in the binary
and not in a nebula around it.  Another important feature of the flare
spectra is that lines are slightly broader than in the mean spectra,
indicating that they are formed mainly in the regions of the disc that
rotate faster, i.e the inner regions, whereas the lines that
contribute to the mean spectrum are formed in lower velocity regions.

\subsection{Central spike}

When we plot all the spectra together versus orbital phase we notice
that the central spike behaves like an S-wave with a radial velocity
semiamplitude of about 10\,km\,s$^{-1}$. This S-wave completes a whole
cycle in an orbit indicating that its origin is somewhere in the
binary and not in a nebula around it (as that would produce a
stationary peak). We carried out multi-gaussian fittings to the
profiles of the lines and fitted the velocities associated with the
gaussian corresponding to the central spike by the $\gamma -
V_{X}\cos{2\pi\phi}+V_{Y}\sin{2\pi\phi}$ function. The parameters of
the fit are shown in table~\ref{tab1}. We could not fit
\hbox{$\hbox{He\,{\sc i}\,$\lambda$4388\,\AA}$}\ and
\hbox{$\hbox{He\,{\sc i}\,$\lambda$4922\,\AA}$}\ accurately as the
central spike is double peaked. The radial velocity semiamplitudes
measured range between $\sim$6--12\,km\,s$^{-1}$ similar to the
$\sim$10\,km\,s$^{-1}$ semiamplitude measured by \cite{nrs81} and
\cite{m99}. The systemic velocity $\gamma$ is $\sim$40\,km\,s$^{-1}$
for most lines apart from \hbox{$\hbox{He\,{\sc
      ii}\,$\lambda$4686\,\AA}$}\ with $\sim$20\,km\,s$^{-1}$ and
\hbox{$\hbox{He\,{\sc i}\,$\lambda$5015\,\AA}$}\ with
$\sim$8\,km\,s$^{-1}$.

A systemic velocity of the order of $\sim$35\,km\,s$^{-1}$ could be
the result of gravitational redshift of light emitted by a
$\sim$0.6\,M$\odot$ white dwarf. However, to produce a redshift of
only 8\,km\,s$^{-1}$, the \hbox{$\hbox{He\,{\sc
      i}\,$\lambda$5015\,\AA}$}\ line should form at a radius 4 times
larger than that of the white dwarf, which seems highly unlikely.

\begin{table}
\caption{Velocity parameters of central spike measured after fitting
  the profile of the line with multiple gaussian functions}
\begin{center}
\renewcommand{\arraystretch}{1.4}
\setlength\tabcolsep{5pt}
\begin{tabular}{llllll}
\hline\noalign{\smallskip}
Line & $\gamma$ & $V_{X}$ & $V_{Y}$\\
     & km\,s$^{-1}$ & km\,s$^{-1}$ &km\,s$^{-1}$\\
\noalign{\smallskip}
\hline
\noalign{\smallskip}
\hbox{$\hbox{He\,{\sc i}\,$\lambda$4388\,\AA}$}\ & -- & -- & --\\
\hbox{$\hbox{He\,{\sc i}\,$\lambda$4471\,\AA}$}\ & 49.02 $\pm$ 0.18 & 6.24 $\pm$ 0.27 & 0.39 $\pm$ 0.06\\
\hbox{$\hbox{He\,{\sc ii}\,$\lambda$4686\,\AA}$}\ & 20.42 $\pm$ 0.27 & 10.46 $\pm$ 0.40 & 0.91 $\pm$ 0.09\\
\hbox{$\hbox{He\,{\sc i}\,$\lambda$4713\,\AA}$}\ & 38.25 $\pm$ 0.29 & 11.46 $\pm$ 0.43 & 1.84 $\pm$ 0.13\\
\hbox{$\hbox{He\,{\sc i}\,$\lambda$4922\,\AA}$}\ & -- & -- & --\\
\hbox{$\hbox{He\,{\sc i}\,$\lambda$5015\,\AA}$}\ & 7.64 $\pm$ 0.33 & 8.55 $\pm$ 0.47 & 0.89 $\pm$ 0.12\\
\hbox{$\hbox{He\,{\sc i}\,$\lambda$6678\,\AA}$}\ & 42.98 $\pm$ 0.23 & 10.33 $\pm$ 0.35 & -0.67 $\pm$ 0.07\\
\hbox{$\hbox{He\,{\sc i}\,$\lambda$7065\,\AA}$}\ & 46.20 $\pm$ 0.24 & 6.56 $\pm$ 0.35 & -1.10 $\pm$ 0.11\\
\hbox{$\hbox{He\,{\sc i}\,$\lambda$7281\,\AA}$}\ & 42.59 $\pm$ 0.43 & 10.70 $\pm$ 0.63 & -0.35 $\pm$ 0.11\\
\hline
\end{tabular}
\end{center}
\label{tab1}
\end{table}

\section{Doppler maps}

\begin{figure}[t]
\begin{center}
\includegraphics[angle=-90,width=.97\textwidth]{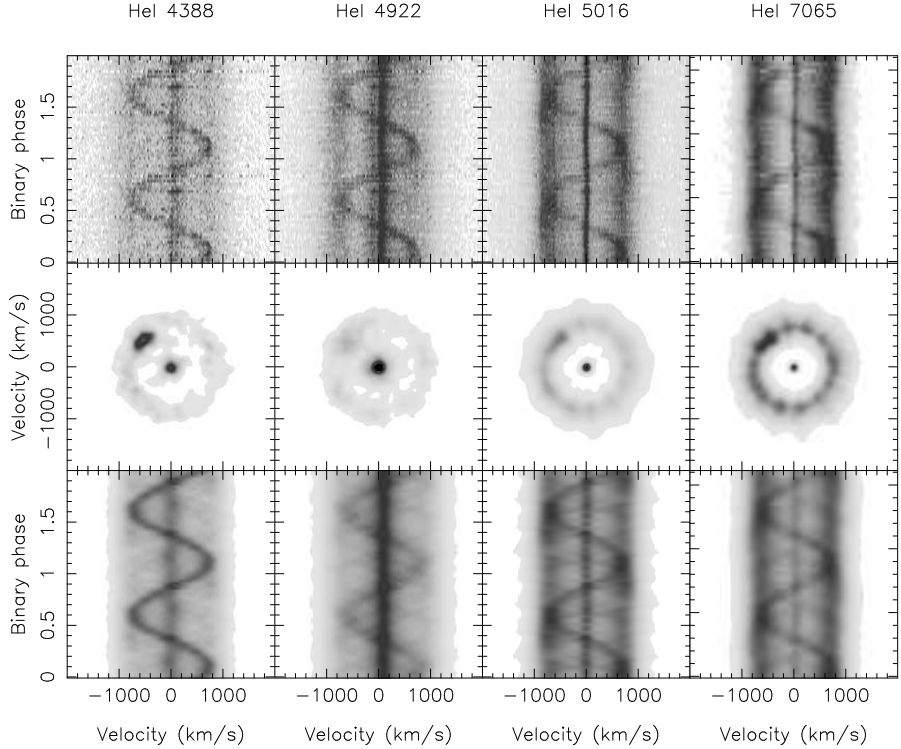}
\end{center}
\caption[]{Trail spectra of 4 helium emission lines ({\bf top}),
  Doppler maps calculated from the spectra ({\bf middle}) and spectra
  computed back from the maps ({\bf bottom})}
\label{eps3}
\end{figure}

We modified the orbital ephemeris given by \cite{m99} so the
modulation of the central spike corresponded to the motion of the
white dwarf in the binary. Using the maximum entropy technique we
obtained Doppler maps for the emission lines and show them, for 4 of
the lines, in Fig.~\ref{eps3}. For maps of all the lines see
\cite{mmwn00}. The top panels present the spectra binned in orbital
phase and plotted twice. The central spike is the strongest feature in
almost all the lines. In some cases it appears to be double-peaked
(\hbox{$\hbox{He\,{\sc i}\,$\lambda$4388\,\AA}$}\ and
\hbox{$\hbox{He\,{\sc i}\,$\lambda$4922\,\AA}$}) which we suggest is
the result of Stark broadening.  The Stark effect does not affect
helium lines in the same way as hydrogen lines. In hydrogen the effect
is symmetrical whereas in helium it results in some forbidden
transitions being allowed. What we believe we are seeing is two of
those forbidden transitions which happen to lie very close to He\,{\sc
  i} $\lambda$4388\,\AA\ and $\lambda$4922\,\AA, see
\cite{bw98,bwb95,bwb97} for details. This indicates that the central
spike must form in a high electron density region, which favours the
white dwarf as its origin.

Also clear in the trails are the two peaks equidistant from the rest
wavelength that correspond to the accretion disc, and a sinusoidal
component that moves between the two peaks. The middle panels are the
Doppler tomograms obtained using MEMSYS techniques \cite{mh88}. The
emission in the centre of the map, therefore at low velocities,
corresponds to the central spike. The red and blue peaks of the lines
map into a ring around the centre of mass that corresponds to the
accretion disc. The sinusoidal component maps into an emission region
on the top left quadrant of the map. This is the position in the map
where we expect to find any emission coming from the bright spot. We
observe that the bright spot shows a complex structure, stretched
along the accretion disc, for some lines. This behaviour had been
observed previously \cite{m99}. When Marsh \cite{m99} measured the
radial velocity of the spot at different orbital phases, he realised
that it moved in a semi-sinusoidal fashion, the values of the velocity
always being between the radial velocities of the stream and the disc
at the bright spot position. The bottom panels of Fig.~\ref{eps3}
present the trails of the spectra computed back from the Doppler maps.
The agreement between the real and computed data assures us that the
structures seen in the maps are real and not artifacts.

One last thing to notice in the \hbox{$\hbox{He\,{\sc
      i}\,$\lambda$5015\,\AA}$}\ and \hbox{$\hbox{He\,{\sc
      i}\,$\lambda$7065\,\AA}$}\ maps is that the disc appears to be
slightly elliptical. The reason for this is not clear to us.

\section{A few puzzles still}

Although we can explain some of the behaviour observed in GP~Com,
there are still several peculiarities we do not fully understand. Why
is the central spike redshifted by different amounts for different
lines? We notice in some of the maps that the accretion disc seems
elliptical: is that real? And if so, why? Why do only the inner
regions of the disc contribute to the flaring? What is causing the
flaring? What is causing the long term variability observed on this
source? What does the fact that Stark broadening affects the lines so
much indicate?

%INDEX%%%%%%%%%%%%%%%%%%%%%%%%%%%%%%%%%%%%%%%%%%%%%%%%%%%%%%%%%%%%%%%
% Please check with the editor of your book whether he plans to
% include a "mutual" subject index - if so, please code your entries
% in the standard syntax. For your own purposes you may print your
% "personal" index by using the following commands:
%
%\clearpage
%\addcontentsline{toc}{section}{Index}
%\flushbottom
%\printindex
%%%%%%%%%%%%%%%%%%%%%%%%%%%%%%%%%%%%%%%%%%%%%%%%%%%%%%%%%%%%%%%%%%%%%

\end{document}